\documentclass{elsart}
\usepackage{amsmath}

\usepackage{graphics}
\usepackage{graphicx}
\usepackage{amssymb,epsfig,picinpar,graphics,float,amssymb,verbatim}

\usepackage{amssymb}

\begin{document}

\begin{frontmatter}



\title{Suppression of selection among broken-symmetry states by cross-correlated noises in double-well system}

\author {Iurii V. Gudyma\corauthref{cor1}}
\ead{yugudyma@gmail.com} \corauth[cor1]{Corresponding author.}
\author {, Artur Iu. Maksymov}
\address {Department of General Physics, Chernivtsi National University,
2 Kotsubinskogo str., 58012 Chernivtsi, Ukraine}

\author{}

\address{}

\begin{abstract}
Suppression of selection among broken-symmetry states has been
found in double-well system resulting from two simultaneous
correlated white noises, one additive and the other
multiplicative. Symmetry-asymmetry-symmetry transition is
re-entrant as a function of the additive noise intensity for small
multiplicative noise.
\end{abstract}

\end{frontmatter}

\section{Introduction}
\label{} The study of nonlinear dynamical systems perturbed
cross-correlated noises has become an attractive subject in recent
years
\cite{Chauhuri08,Den06,Jin05,Luo03,Mei03,Bag02,Git02,Cao00,Jia00}.
The cross-correlated noise processes were first considered by
Fedchenia \cite{Fedchenia88} in the context of hydrodynamics of
vortex flow of fluctuations from a common origin that appear in
the time evolution equation of dimensionless modes of flow rates.
First Fulinski and Telejko discuss a bistable kinetic model under
the simultaneous influence of additive and multiplicative Gaussian
white noises in 1991  \cite{Ful91}. Behavior of dynamical model is
a qualitative change of the state of a system as the degree of
correlation between noises increased. An appropriate formalism and
basic relations for cross-correlated noises were developed in the
pioneering study of Wu et al. \cite{Wu94}. The effect of two
correlated white noises on the giant suppression of the escape
rate in a double-well system has received in Ref. \cite{Mad96}.
Cross-correlation in bistable system can induce re-entrant phase
transition \cite{Li96}. The authors of Ref. \cite{Jia96} have
investigated the steady-state regime of the bistable kinetic
system in the presence of correlations between the noises. The
authors of Ref. \cite{Ke99} proposed a new mechanism for a
noise-induced current by correlated additive and multiplicative
noises under symmetric periodic potentials. In reference
\cite{Tes00}, the idea of correlation between additive and
multiplicative noises has been generalized to the study of
stochastic resonance. We concluded that the noise was found to
change the stability of the metastable system: maximum of bimodal
distribution passes into minimum and vice versa while changing one
of the correlated noises \cite{Gudyma06}. As a result, the
initially asymmetric system can be reduced to symmetric one.

In this work, we focus on simultaneous influence of additive and
multiplicative noises on double-well system. For describing the
stationary properties of the system we should simultaneously
consider the additive thermal noise and the multiplicative
parametric fluctuations.

The paper is organized as follows. In next section, the general
model of the double-well system under cross-correlated noises is
briefly discussed. The model we analyze consist of a system with
symmetric potential and small asymmetric potential component. In
section\ \ref{sec:stationary} we present derivation of stationary
probability distribution function in the Gibbs form. In section\
\ref{sec:simulation} we show the results obtained by numerical
integrating the dynamical stochastic equations. The final section
summarizes the main results of the paper.
\section{Model}
\begin{figure}[h]
\center \epsfig{file=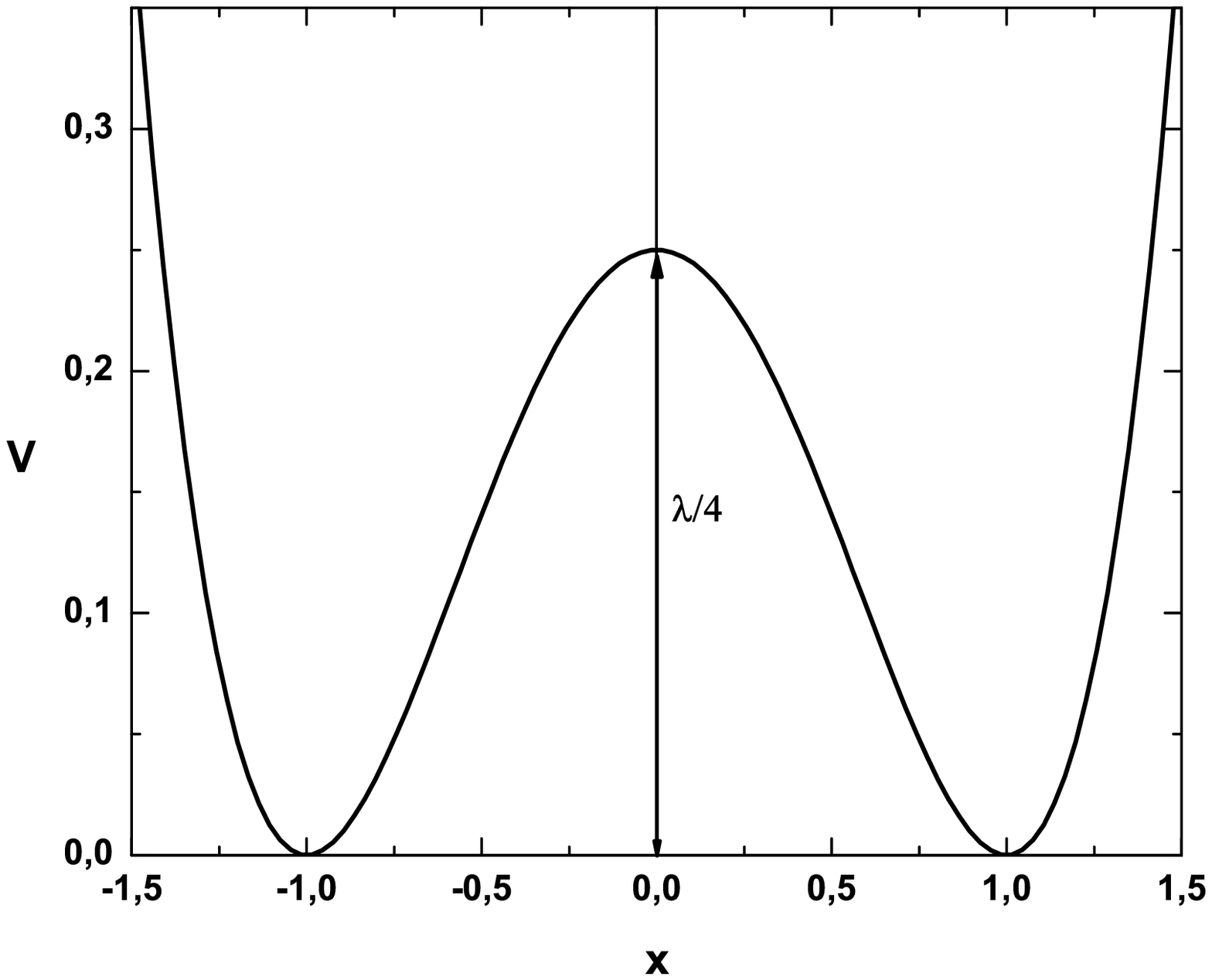, width=.45\textwidth}
\epsfig{file=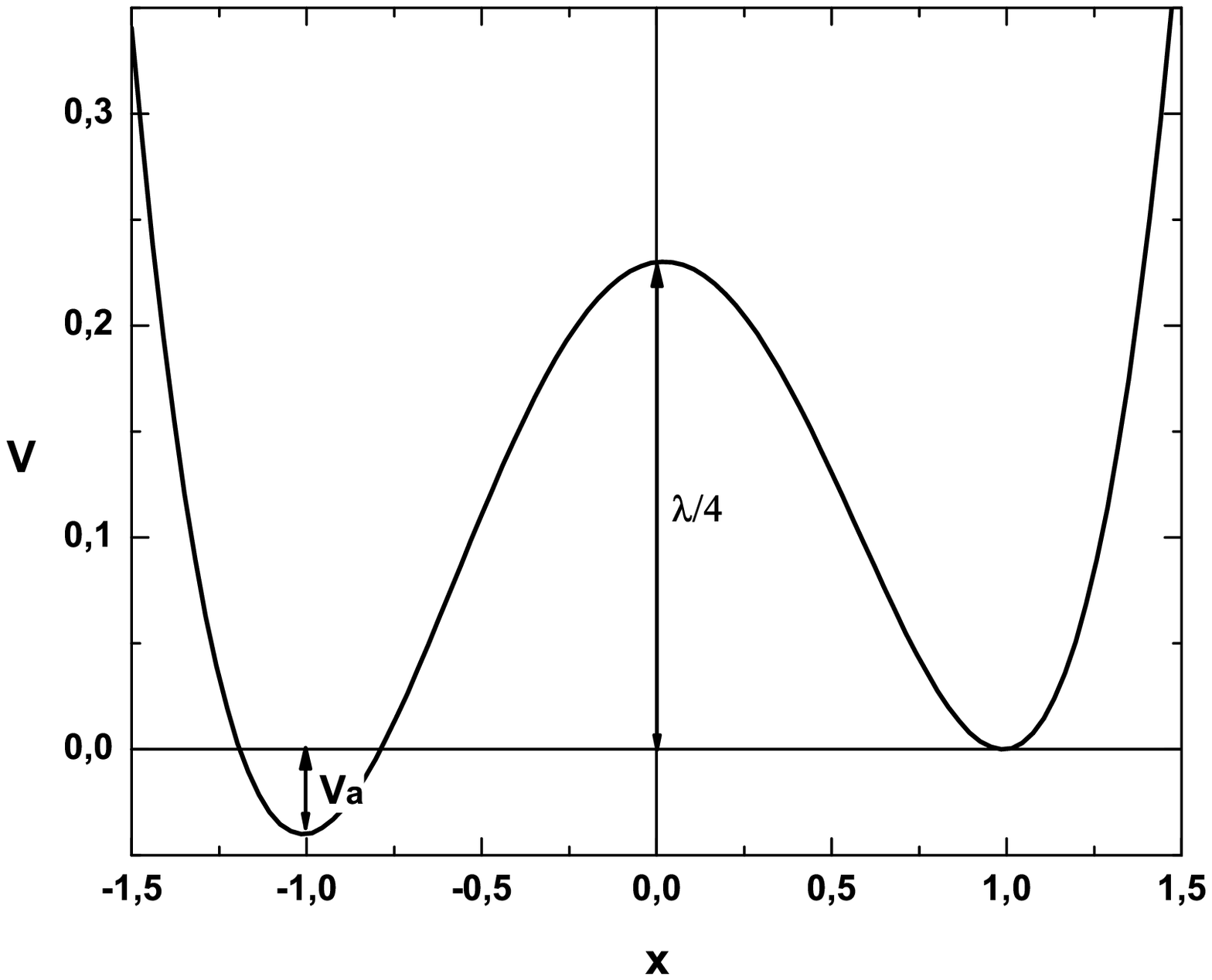, width=.45\textwidth} \caption{Various
types of bistable potentials of the form given by
Eq.~(\ref{potential}) for $\lambda=1$. (a) symmetric potential
  $\Delta=0$; (b) asymmetric potential
  $\Delta=1$.\label{fig:potential}}
\end{figure}

We consider a bistable system with potential consisting of a
symmetric double-well term
\begin{align}
V_{s}(x)=\frac{\lambda}{4}\left(x^2-a^2\right)^2  \label{sym}%
\end{align}
and an additional small asymmetric term:
\begin{align}
V_{a}(x)=\frac{\Delta}{a}\left(x-a\right).  \label{asym}%
\end{align}
The parameter $\Delta$ fixes asymmetry of the potential. It is a
small interaction or bias that selects among the broken-symmetry
states. The full potential is given by
\begin{align}
V (x)=V_{s}(x)+V_{a}(x).  \label{potential}%
\end{align}
We consider a potential of the type depicted in
Fig.~\ref{fig:potential}, with a metastable minimum at $x_{+}=+1$
and a stable minimum of $x_{-}=-1$. Thereto value of $a=1$ were
used for computation hereinafter in this paper.
Note, such weakly asymmetric case, when right well is slightly
higher than left one, corresponds smaller density of states in the
right well than in the left one. Following Callan and Coleman
\cite{Callan77} we call the phase corresponding to the minimum at
position $x_{+}$ the false vacuum and the one corresponding to the
minimum at position $x_{-}$ the true vacuum. Callan and Coleman
presented an approach for analyzing the decay of an unstable
vacuum in the framework of Euclidean quantum field theory.

For the system coupled to noisy reservoir the relevant Langevin
equation of motion can be written as
\begin{align}
\frac{\partial x }{\partial t}  =-\frac{\partial V }{\partial x}+\varepsilon\left(  t\right). \label{general}%
\end{align}
The first term corresponds to the dissipative term as well as the
external applied force. The second term in Eq. (\ref{general})
refers to an external additive noise. We assume that barier
potential $\lambda$ fluctuates
\begin{align}
V (x)  =\frac{\lambda\left(  t\right)}{4}\left(  x^2-a^2\right)^2+\frac{\Delta}{a}\left( x-a\right), \lambda\left(  t\right)=\lambda+\xi\left(  t\right).\label{concrete}%
\end{align}
In general, we express the thermal fluctuation of the system as
additive noise and the effect of the external environmental
fluctuation on the system as multiplicative noise. A model with a
fluctuating barrier and an additive noise was used in Ref.
\cite{FedcheniaUsova}. Linear fluctuations of the potential
barrier were considered uncorrelated or correlated to the additive
noise. The noises $\varepsilon\left( t\right)$ and $\xi\left(
t\right)$ denote white Gaussian noises with zero mean and
correlation function
\begin{equation}
\langle \varepsilon\left( t\right) \varepsilon\left( t+\tau\right)
\rangle=2\epsilon^2\delta\left( \tau\right), \label{varepsilon}%
\end{equation}
\begin{equation}
\langle \xi\left( t\right) \xi\left( t+\tau\right)
\rangle=2\sigma^2\delta\left( \tau\right). \label{xi}%
\end{equation}
Hence the equation of motion can be obtained from Eq.
(\ref{general}) with using expression (\ref{concrete})
\begin{align}
\frac{\partial x }{\partial t}  =-\lambda x \left(  x^2-a^2\right) - \frac{\Delta}{a} - x \left(  x^2-a^2\right)\xi\left( t\right)+\varepsilon\left(  t\right). \label{explicit}%
\end{align}
Below we assume Eq. (\ref{explicit}) to be the Stratonovich
stochastic differential equation with the supplementary condition
\begin{equation}
\langle  \varepsilon\left( t\right)\xi\left( t+\tau\right)
\rangle=\langle  \xi\left( t\right)\varepsilon\left( t+\tau\right)
\rangle=2\chi\epsilon\sigma\delta\left( \tau\right). \label{cross}%
\end{equation}
Value $\chi$ denotes the degree of correlation between additive
$\varepsilon\left( t\right)$ and multiplicative $\xi\left(
t\right)$ noises. The correlation between the multiplicative and
additive noises realizes the hierarchical coupling of the system
to the heat bath. Here $\epsilon$ and $\sigma$ are the strength of
noises $\varepsilon\left( t\right)$ and $\xi\left( t\right)$,
respectively.
%
\begin{section}{Stationary probability distribution function}\label{sec:stationary}
From equation (\ref{explicit}) with conditions (\ref{varepsilon}),
(\ref{xi}) we can get the stochastic equivalent Langevin equation
with one white-noise term \cite{Wu94}
\begin{align}
\frac{\partial x }{\partial t}  =\Phi\left(  x\right)+\Gamma\left(  x\right)\gamma\left(  t\right), \label{Wu}%
\end{align}
where $\Phi\left(  x\right)=-\lambda x \left(
x^2-a^2\right)-\Delta/a$, and $\gamma\left(  t\right)$ is a
Gaussian white noise having zero mean magnitude and a delta-like
correlation function
\begin{equation}
\langle \gamma\left( t\right) \gamma\left( t+\tau\right)
\rangle=2\delta\left( \tau\right). \label{gamma}%
\end{equation}
The expression for value $\Gamma\left(  x\right)$ is determined by
the following simple procedure: Let the correlation of
$\Gamma\left( x\right)\gamma\left(  t\right)$ in Eq.~(\ref{Wu}) be
equal to the correlation of $- x \left( x^2-a^2\right)\xi\left(
t\right)+\varepsilon\left(  t\right)$ in Eq.~(\ref{explicit}).
Following this approach, one can see that the required expression
is
\begin{equation}
\Gamma\left( x\right) = \left[ \sigma^2 \varphi^2\left( x \right)+2\chi\epsilon\sigma \varphi\left( x \right)+\epsilon^2\right]^{1/2}, \label{Gamma}%
\end{equation}
where
\begin{equation}
\varphi\left( x \right) = -x\left( x^2-a^2 \right). \label{varphi}%
\end{equation}
The equation~(\ref{Wu}) to be understood in the Stratonovich
interpretation.

The Fokker-Plank equation corresponding to the stochastic
differential equation (\ref{Wu}) reads
\begin{align}
\frac{\partial P \left(  x, t\right) }{\partial t}  =-\frac{\partial}{\partial x}\left\{\Phi\left(  x\right)+\frac{1}{2}\frac{\partial}{\partial x}\left[\Gamma\left(  x\right)\frac{\partial}{\partial x}\Gamma\left(  x\right)\right] \right\} P \left(  x, t\right). \label{FPE}%
\end{align}
This equation governs the time evolution of the system's
probability distribution $P\left(  x, t\right)$   at time $t$. In
the case of natural or instantly reflected boundary condition no
probability current exists. The stationary probability
distribution being
\begin{align}
 P_{st} \left( x\right) =N\cdot\exp \left\{\int_0^x du\frac{\Phi\left(  u\right)-\Gamma\left(  u\right)\Gamma'\left(  u\right)}{\Gamma^2\left(  u\right)}\right\}, \label{stat_prob}%
\end{align}
where $N$ denotes a normalization constant. The simplest way to
analyze the stochastic dynamics of a system expressed by
Eq.(\ref{Wu}) is to take the Gibbs form of the stationary
probability distribution function
\begin{align}
 P_{st} \left( x\right) \sim \exp \left\{-V_{eff}\left( x\right)\right\}. \label{stat_prob}%
\end{align}
The extrema of the effective potential $V_{eff}$ correspond to the
stationary fixed points of the noise-sustained dynamics. For a
bistable system, the probability distribution has two maxima which
correspond to the stochastic steady states. Noted that the
stationary probability distribution exhibits a symmetric bimodal
structure for $\Delta=0$ and $\chi=0$.
\end{section}
\begin{section}{Numerical simulation}\label{sec:simulation}

We numerically simulated the double-well stochastic system described
by Eq. (\ref{explicit}). For that we employed the second-order
stochastic Runge-Kutta type method (Heun's algorithm) for
integrating the dynamical equations of motion \cite{Miguel} with an
integration time step $\Delta t=10^{-3}$. Gaussian white noise is
generated using the Box-Muller algorithm. All calculated quantities
are dimensionless.

Fig.\ref{fig:reentrant} displays the phase diagram of the model in
the $\chi-\varepsilon$ and $\chi-\sigma$ parameter planes for
fixed parameters $\sigma$ and $\varepsilon$ respectively.

\begin{figure}[h]
\center \epsfig{file=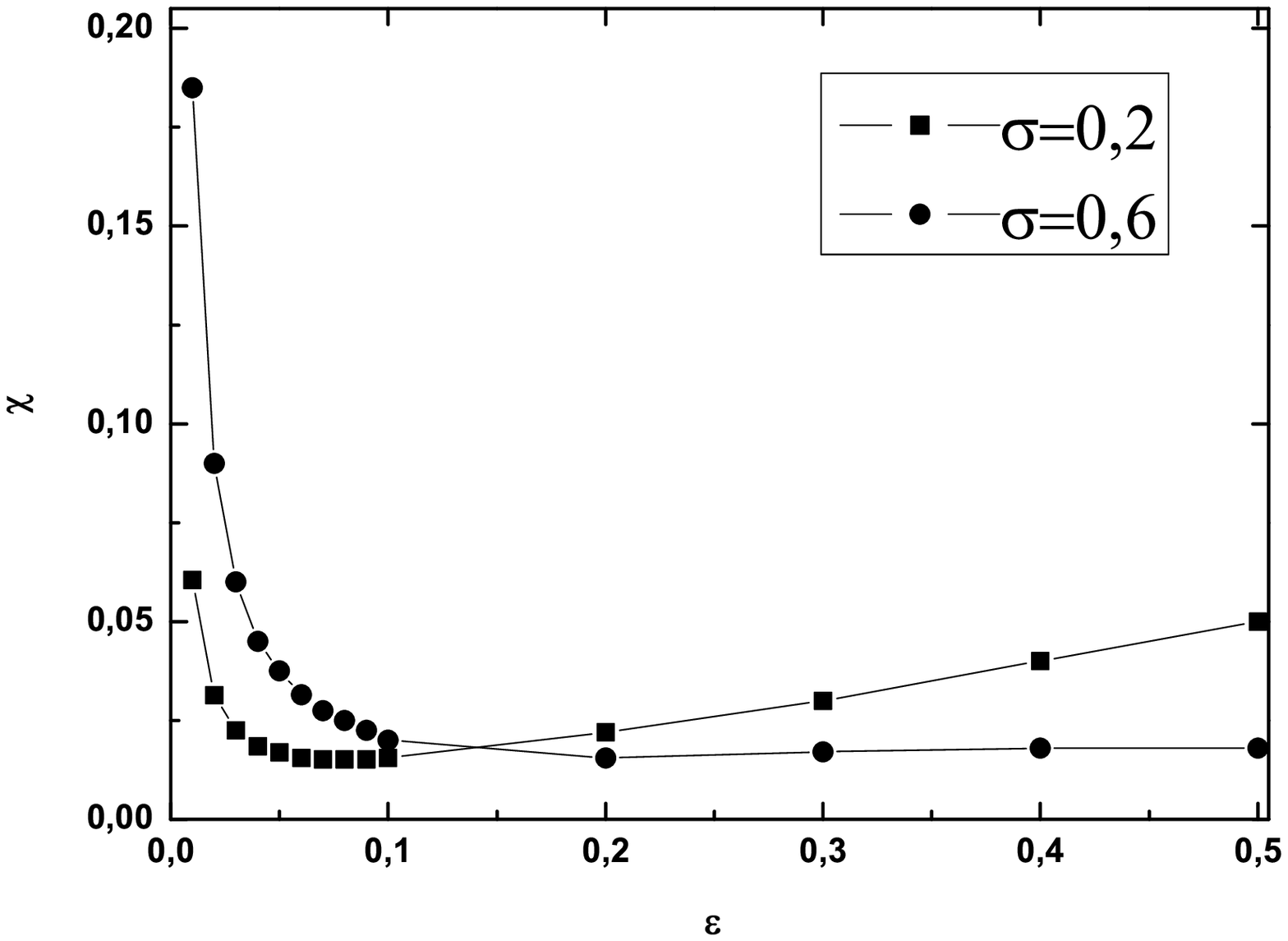, width=.45\textwidth}
\epsfig{file=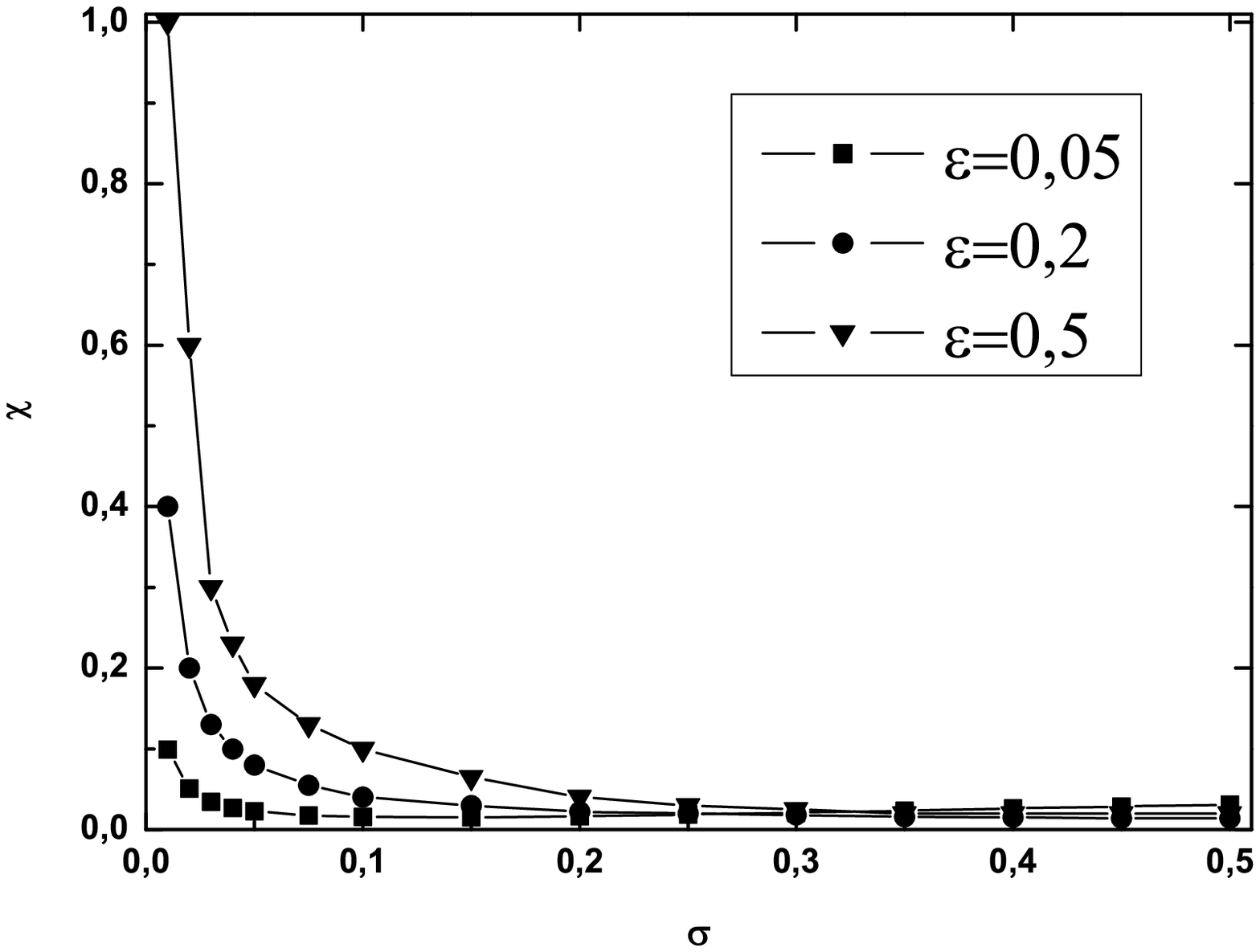, width=.45\textwidth} \caption{The phase
diagram. (a) The plot of the degree of correlation $\chi$ between
the multiplicative and additive noises \emph{vs} the additive noise
intensity $\varepsilon$ for fixed values of $\sigma$. (b) The plot
of the degree of correlation $\chi$ between the multiplicative and
additive noises \emph{vs} the multiplicative noise intensity
$\sigma$ for fixed values of $\varepsilon$.\label{fig:reentrant}}
\end{figure}

The region with left maximum peak and right maximum peak are
separated by a line of equal probability. On this line the
potential $V_{eff}\left( x\right)$ is formal invariant under
$x\rightarrow-x$ while the function $P_{st} \left( x\right)$ is a
symmetric function of $x$. Thus noise suppressed of selection
among broken-symmetry states. For $\sigma$ not too large the
increase of the additive noise intensity causes the re-entrant (
symmetric-asymmetric-symmetric probability) transition.

In order to eliminate the influence of mutually correlated noises
on the double-well system in Fig.\ref{fig:barrier} we show the
same curves in the $\chi-\lambda$ parameter planes. This
representation clear indicate that for deep potential low degree
of correlation between noises is quite enough for suppression of
selection among broken-symmetry states. Moreover, the influence of
additive noise $\varepsilon$ decreases with increase of $\lambda$.
\begin{figure}[h]
\center \epsfig{file=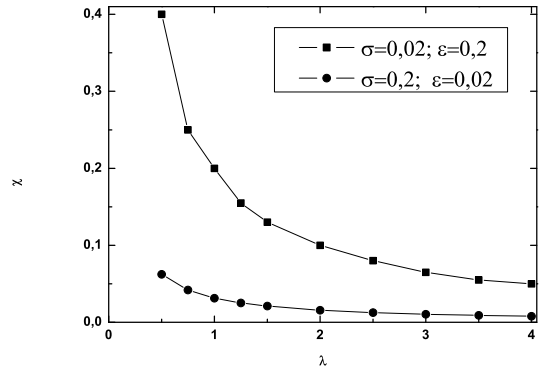,width=.6\textwidth} \caption{Plot
of $\chi$ \emph{vs} $\lambda$ for fixed values of $\sigma$ and
$\varepsilon$. \label{fig:barrier}}
\end{figure}

It must be emphasized that the lines in figures
\ref{fig:reentrant} to \ref{fig:barrier} correspond to stationary
probability distribution function with symmetric maxima and
separate two sectors corresponding a spontaneous breakdown of
parity. As already mentioned, existence of cross-correlation
between additive and multiplicative noises may lead to
symmetrization of the stationary probability distribution. For
$\chi=0$ and $\Delta\neq0$ the shape of distribution is
asymmetric.

\end{section}
\begin{section}{Discussion}\label{sec:discuss}

In this paper, we have analyzed the behavior of a double-well
system under cross-correlated noises. In part, we have studied
numerically for the lines corresponding to equal probability of
maxima of this bistable system. We have shown that the
cross-correlated noises suppressed of selection among
broken-symmetry states. For $\sigma$ not too large this effect is
re-entrant as a function of $\varepsilon$. Only the combined
actions of the multiplicative noise and the additive noise causes
such behavior. When the noises are uncorrelated, the stationary
probability distribution exhibits an asymmetric bimodal structure
for $\Delta\neq0$ and a symmetric bimodal structure for
$\Delta=0$.

\end{section}


\end{document}